\documentclass[12pt]{article}
\usepackage{epsfig}

\begin{document}

\title{Evaluation of the escape widths of the giant dipole resonances in the fermi-liquid theory}
\author{V. A. Sadovnikova,\\
Petersburg Nuclear Physics Institute,  NRS Kurchatov Institute,  \\
Gatchina, St.~Petersburg 188300,
Russia}
\date{\today}
\maketitle

\begin{abstract}
In the theory of the finite fermi-systems \cite{Mi}, it was shown that giant resonances in nuclei can be consider as the zero-sound excitations which exhaust the large part of the energy-weighted sum rules.
In the framework of  \cite{Mi} the solutions of the zero-sound dispersion equation in the symmetric nuclear matter, $\omega_s(k)$, are considered.
 The method of calculation of these solutions is based on the analytical structure of the polarization operators $\Pi(\omega,k)$.
The solutions of the dispersion equation, which are real at small $k$, become complex with $k$ increasing  when the overlapping of the collective and $1p1h$ modes starts.
The imaginary part of $\omega_s(k)$ is the result of the collective zero-sound excitation decay to the real particle-hole pairs and can be compared with the escape width of resonances.
We  compare the experimental energy and  escape width of the giant dipole  resonance (GDR) in the nucleus  A  with ${\rm Re}\,\omega_s(k)$ and ${\rm Im}\,\omega_s(k)$ taken at a definite wave vector $k=k_A$.
\end{abstract}

\section {Introduction}
In the theory of the finite fermi systems \cite{Mi} it was shown that the giant resonances in  nuclei can be considered as the zero-sound excitations which exhaust   the large part of the energy-weighted sum rules \cite{MZL,SW}. The same approach is used in the random phase approximation  \cite{SW,I31,Ur}, in the gas-liquid model \cite{De2}.
In this paper the attempt is made to obtain the information about the giant resonances in nuclei on basis of the solutions of the dispersion equation for the zero-sound excitations in the nuclear matter.

We consider the zero-sound dispersion equation in the symmetric nuclear matter
\begin{equation}\label{1}
1 =  C_0 F\Pi(\omega,k),
\end{equation}
here $C_0=N_0^{-1}$,  $N_0=2p_Fm/\pi^2$ is the state density of two sorts of nucleon on the Fermi-surface, $p_F$ is the Fermi momentum. The parameter $F$ is the constant of the Landau-Migdal effective quasiparticle-quasihole interaction \cite{Mi}.

\begin{equation}\label{2}
{\cal F}(\vec\sigma_1,\vec \tau_1;\vec\sigma_2,\vec \tau_2) = C_0\left(F + F'(\vec\tau_1\vec\tau_2) + G(\vec\sigma_1\vec\sigma_2)
 + G'(\vec\tau_1\vec\tau_2)\, (\vec\sigma_1\vec\sigma_2)\right),
\end{equation}
here $\vec\sigma$, $\vec \tau$  are Pauli matrices in the spin and isospin space. The polarization operator $\Pi(\omega,k)$  is the integral over a nucleon particle-hole loop. Here the simple model of the interaction ${\cal F}(\vec\sigma_1,\vec \tau_1;\vec\sigma_2,\vec \tau_2)$ is used when the interaction is determined by the set of constants. In such model the equation (\ref{1}) is true for  all channels of the interaction: scalar ($F$), isospin ($F'$), spin ($G$) and spin-isospin ($G'$). But the values of $F,F',G,G'$ are different \cite{Mi}. For example, the constant, responsible for the monopole isoscalar excitations is very small in the nuclear matter. The polarization operator is taken in a usual way:
\begin{equation}\label{3}
\Pi(\omega,k) = -4\Phi(\omega,k) = -4 (\phi(\omega,k) + \phi(-\omega,k))\, .
\end{equation}
Functions $\phi(\omega,k)$ are the Migdal's function  \cite{Mi}
\begin{equation} \label{4}
\phi(\omega,k) =\ \frac1{4\pi^2}\
\frac{m^3}{k^3}
\left[\frac{a^2-b^2}2 \ln\left(\frac{a+b}{a-b}\right)-ab\right] ,
\end{equation}
here $a=\omega-(k^2/2m)$, $b=kp_F/m$. The equation
 (\ref{1}) can be written in the form
\begin{equation}\label{1a}
1 + 4C_0F\Phi=0.
\end{equation}

The method of calculation of (\ref{1}) is presented in \cite{2,3}. The polarization operator $\Pi(\omega,k)$ (\ref{3},\ref{4})  has two overlapping cuts on the real axis of the complex $\omega$-plane. These are the logarithmic cuts  of the functions $\phi(\omega,k)$ and $\phi(-\omega,k)$.  The physical sense  of the cut  is the appearance of the real  $ph$-pair  at the definite values of $F$, $p_F$, $m$, $k$ in the particle-hole loop in $\Pi(\omega,k)$.
For not too large wave vectors $k$ the collective excitations have the  frequencies which are larger than the $ph$-pair ones. It means that $\omega_s(k)$ lies on the real axis of the complex $\omega$-plane more to the right than the cut. With increasing $k$, the cut ($ph$ mode) overlaps with the solution $\omega_s(k)$ (collective mode). The solutions $\omega_s(k)$ become complex and go through the cut to the nearest unphysical sheet of the Riemann surface of the  logarithm in (\ref{4}). The physical nature of the imaginary part of $\omega_s(k)$ is determined by the structure of the cut, i.e. the excitation of the real  $ph$-pair with the further escape of a particle. So we accept that ${\rm Im}\,\omega_s(k)$ corresponds to  the escape width of the excitations.

It was shown in \cite{PN} that the zero-sound excitations in the matter exhaust  the large part of the energy-weighted sum rules. The giant resonances have the same property by the definition. Besides, the equations of the random phase approximation which are used in calculation of the energies and the width of the giant resonances are  similar to equations for the zero-sound in the nuclei \cite{Mi,Sp}. Therefore it is interesting to see what
can we know about  the giant resonances in the nuclei if to use  $\omega_s(k)$ obtained in the nuclear matter. To establish such relation we must know  the wave vectors $k_A$ which corresponds to the  definite giant  resonance in the nucleus with the atomic number $A$. There are  some models
which consider the nuclei as a drop of a liquid and calculate the energies, $k_A$ and other parameters of resonances \cite{De2,SJ,BV}.

We calculate $\omega_s(k_A)$ for different  $A$ and  compare the real part ${\rm Re}\,\omega_s(k_A)$ with the energies  of the giant resonances. To evaluate the escape width we use $|{\rm Im}\,\omega_s(k_A)|$ which corresponds to the half of the escape width.

The branch $\omega_s(k_A)$ depends on the nuclear matter parameters: the Fermi momentum $p_F$,  the value of  the  effective isovector coupling constant $F'$, the value of the effective mass $m$. Variation of these parameters in the acceptable limits permits us to obtain the reasonable correspondence between the experimental data and $\omega_s(k_A)$.

It should be mentioned that the dispersion law of the zero sound is the linear one and $\omega_s(k_A)$ grows with $k_A$.  The imaginary part of $\omega_s(k_A)$ appears when the overlapping of the collective and the $1p1h$ modes starts (i.e. at a definite $k=k_t$) and then grows by the absolute value with $k_A$ \cite{3,FW}.
In the model \cite{SJ} $k_A$ is inversely proportional to  $A^{1/3}$;  $k_A$ (and, consequently, ${\rm Re}\,\omega_s(k)$ and $|{\rm Im}\,\omega_s(k)|$) decrease with $A$. Then the energies and escape widths of the giant resonances obtained on  basis of $\omega_s(k_A)$ decrease  with $A$ increasing.

The approach used in this paper is too simple  to pretend on the satisfactory  description of the giant resonances in  nuclei. But it may be very useful if we need the qualitative dependence of the resonance parameters on such variables as  the temperature, the difference of the proton and neutron densities, on the form of the effective interaction, on the additional open $1p1h$ channels (isobar-particle - nucleon-hole channel, for example).

The essential problem of this approach is the definition of $k_A$ for different excitations. The Steinwedel-Jensen model \cite{SJ,BV} defines $k_A$ for the excitations of the different multipolarity. We follow the paper \cite{BV,BV1} for the giant dipole resonances: $k_A=\frac{\pi}{2R}$, where $R=r_0 A^{1/3}$, $r_0=1.2$~fm. Using relation $k_A$ and $A$ we obtain $\omega_s(A)$. This model \cite{SJ} is good enough for the  giant dipole resonances but it gives too large $k_A$ for the excitations with other multipolarity.

To define the escape widths more  precisely we can make use of the following consideration. The linear dispersion law for the zero-sound $\omega_s(k)=\,c k$ gives us the coefficient of the proportionality $c$. This coefficient  depends on the parameters of the nuclear matter $c=c(F,p_F,m)$ which were used to calculate $\omega_s(k)$. Then, using the phenomenological fit (\ref{17}) $E_{GDR}(A)$, we define $k_A=E_{GDR}(A)/c$ for the nucleus $A$. ${\rm Re}\,\omega_s(k_A)$  is equal to $E_{GDR}(A)$ by the definition and the escape width is determined through ${\rm Im}\,\omega_s(k_A)$.

\section{Solutions of the dispersion equation}
The theory of the finite fermi-systems \cite{Mi} describes the excitations of the dipole collective states with the help of the residual isospin effective quasiparticle interaction $F'$. The values of $F'$ are determined from the experimental data and changed from 1.0 to 2.0 \cite{Mi,GLZ}.

In Fig.~1  the solutions of Eq.~(\ref{1a}) are shown. The real parts  ${\rm Re}\,\omega_s(k)$ are placed  at the positive frequencies
and ${\rm Im}\,\omega_s(k)$ are shown  at the negative frequencies. In Fig.~1 the solutions are presented for the different values of the effective constants in Eq.~(\ref{2}), the nuclear matter densities and effective mass $m$.  The {\it main} set of the nuclear matter  parameters is the following:  $p_F=p_0=268$~MeV, $m=0.8m_0$ ($m_0=940$~MeV), $F$=2 \cite{Mi,Sp}.  The curve $2$ on the Fig.~1\,a,b,c is the same and corresponds to the these parameters.

In the Fig.~1a the branches $\omega_s(k)$ are presented for $F$=0.2,\,1.0,\,2.0. In \cite{FW} it is shown that the frequencies of the zero-sound excitations  grow with $F$ at the fixed $k$ (Fig.~16.1). Our calculations demonstrate and it is seen in Fig.~1a that such behavior takes place for the real  $\omega_s(k)$.  In the region of overlapping of the collective and $ph$ modes the behavior is changed: increasing of $F$ gives the decreasing of ${\rm Re}\,\omega_s(k)$ and increasing of ${\rm Im}\,\omega_s(k)$.

In Fig.~1b the branches $\omega_s(k)$ are shown for different $p_F$. Increasing $p_F$ from 200\,MeV to 300\,MeV results in the growth both ${\rm Re}\,\omega_s(k)$ and ${\rm Im}\,\omega_s(k)$. In Fig.~1c  we demonstrate the branches when $m$ is changed from 600\,MeV to 940\,MeV. Increasing of $m$ gives decreasing of ${\rm Re}\,\omega_s(k)$ and increasing of ${\rm Im}\,\omega_s(k)$.  Thus we have a different dependence on the nuclear matter parameters.

\section{Results for GDR}
To pass from zero-sound excitations in the nuclear matter to the giant resonances in the nuclei we must know the wave vector, corresponding to the definite excitations in a nucleus $A$. We use the modification of the model Steinwedel-Jensen, suggested in \cite{BV,BV1}: $k_A=\frac{\pi}{2r_0 A^{1/3}}$. We compare ${\rm Re}\,\omega_s(A)$ with the phenomenological formula for GDR \cite{Atomic}:

\begin{equation}\label{17}
E_{GDR}(A) = 31.2A^{-1/3} + 20.6 A^{-1/6} \mbox{MeV}.
\end{equation}
In Fig.~2a the curve $1$ is described by (\ref{17}).  The calculation with the main set of parameters  results in too high energies (curve $2$, Fig.~2a) and small width of GDR (curve $2$, Fig.~2b). It is possible to find such parameters of the nuclear matter that the  description of the curve $1$ in Fig.~2a will be satisfactory.
For the heavy nuclei we can choose $p_F=260$~MeV, $F$=1.0  and $m=m_0$ (the curves $4$ in Fig.~2a,b). For the light nuclei we should suppose rather small density in nuclei $p_F=200$~MeV, $F$=1.2 and $m=m_0$ (Fig.~2a,b, curves $3$).

For the nuclei $^{12}C$ from (\ref{17}) we have $E_{GDR}=27.2$~MeV, the experimental escape width is $\Gamma^{\uparrow}_{GDR}\approx 3$~MeV. On the Fig.~2 (the curve $3$) we obtain $E_{GDR}=30$~MeV and $\Gamma^{\uparrow}_{GDR}=7.0$~MeV.
For the nuclei $^{208}Pb$ from (\ref{17}) we have $E_{GDR}=13.7$~MeV, the experimental escape width is $\Gamma^{\uparrow}_{GDR}\approx 0.5-2.0$~MeV. On the Fig.~2 (the curve $4$) we obtain $E_{GDR}=13$~MeV and $\Gamma^{\uparrow}_{GDR}=0.17$~MeV. Thus, we have obtain a satisfactory description of GDR in the heavy nuclei, but for the light nuclei we should suppose a very low nuclear density.

Now we try to describe  the escape widths more precisely. The wave vectors $k_A$ related to the giant dipole resonances in the nuclei are rather small $k_A/p_0\leq 0.5$ for $A>12$ \cite{BV,BV1}. As it is shown in Fig.~1 the dependence $\omega_s(k)=\,c k$  can be considered as the linear one. For the set of parameter: $p_F=260$~MeV,  $m=m_0$, $F'=1.0$ we  obtain $c=0.39\pm0.02$. Define $k_A$ using Eq.~(\ref{17}) as  $k_A=E_{GDR}(A)/c$. Then ${\rm Re}\,\omega_s(k_A)$  is equal to $E_{GDR}(A)$ by the definition and the escape width is defined as $\Gamma^{\uparrow}_{GDR}(A)=2\,|{\rm Im}\, \omega_s(k_A)|$. The calculations give $\Gamma^{\uparrow}_{GDR}(^{12}C)=3.0$\,MeV and  $\Gamma^{\uparrow}_{GDR}(^{208}Pb)=0.36$\,MeV. The escape widths of GDR can be approximated by the formula:
\begin{equation}\label{18}
\Gamma^{\uparrow}_{GDR}(A) =  -5.2\,A^{-1/6}  + 14.2\,A^{-1/3}.
\end{equation}

We can do an observation about the monopole giant resonances (GMR). As it was mentioned, Eq.(\ref{1}) describes the excitations in the different channels and the constant $F$ of the effective interaction (\ref{2}) which is responsible of the monopole excitations, is small in the nuclear matter $F\approx 0.2$ \cite{Mi}. In Fig.~1a we see that at $k/p_0<$ 0.4 the real parts of $\omega_s(k)$ depends on the $F$ weakly, but  $|{\rm Im}\,\omega_s(k)|$ increase quickly. Then we can conclude that  in our simple model the energies of GMR and GDR are close in the heavy nuclei \cite{I31} but the escape widths of GMR are appreciably larger.

The paper is partly supported by RFBR (grant 12-02-00158).

\newpage

\section{Figure captions}

\noindent
FIG.~1. Zero-sound branch of solutions $\omega_s(k)$ for the different values of the effective constant $F$ (figure $a$), the Fermi momentum $p_F$ (figure $b$) and the effective mass $m$ (figure $c$).
${\rm Re}\,(\omega_s)$ are drawn at the positive $\omega$, ${\rm Im}\,(\omega_s)$ are shown at the negative $\omega$. The blobs mark the appearance of  ${\rm Im}\,(\omega_s)\neq0$ the for the curves 1,2,3.
Fig.~$a$: $F$=0.2, 1.0, 2.0 (numbers 1,3,2, correspondingly).
Fig.~$b$: $p_F$=200, 268, 300\,MeV (numbers 1,2,3),
Fig.~$c$: $m$=500, 750, 940\,MeV (numbers 1,2,3).

\noindent
FIG.~2.  Comparison of the calculated and experimental (\ref{17}) energies and width of GDR. At Fig.~2a  we compare $E_{GDR}(A)$ (\ref{17}) and ${\rm Re}\,\omega_s(A)$. At Fig.2b ${\rm Im}\,\omega_s(A)$ are shown. The curve $1$ (solid) in Fig.2a corresponds to (\ref{17}) and in Fig.2b it corresponds to   (\ref{18}). The curves $2$ (dash-dotted) present the  solution with the nuclear matter parameters:  $p_F$=268~MeV, $m=0.8m_0$ and $F$=2.0. The curves $3$ (dashed) calculated with the parameters:  $p_F$=200~MeV, $m=m_0$ and $F$=1.2. The dotted curves $4$ is obtained for  $p_F$=260~MeV, $m=m_0$ and $F$=1.0.

\newpage

\begin{figure}
\centering{\epsfig{figure=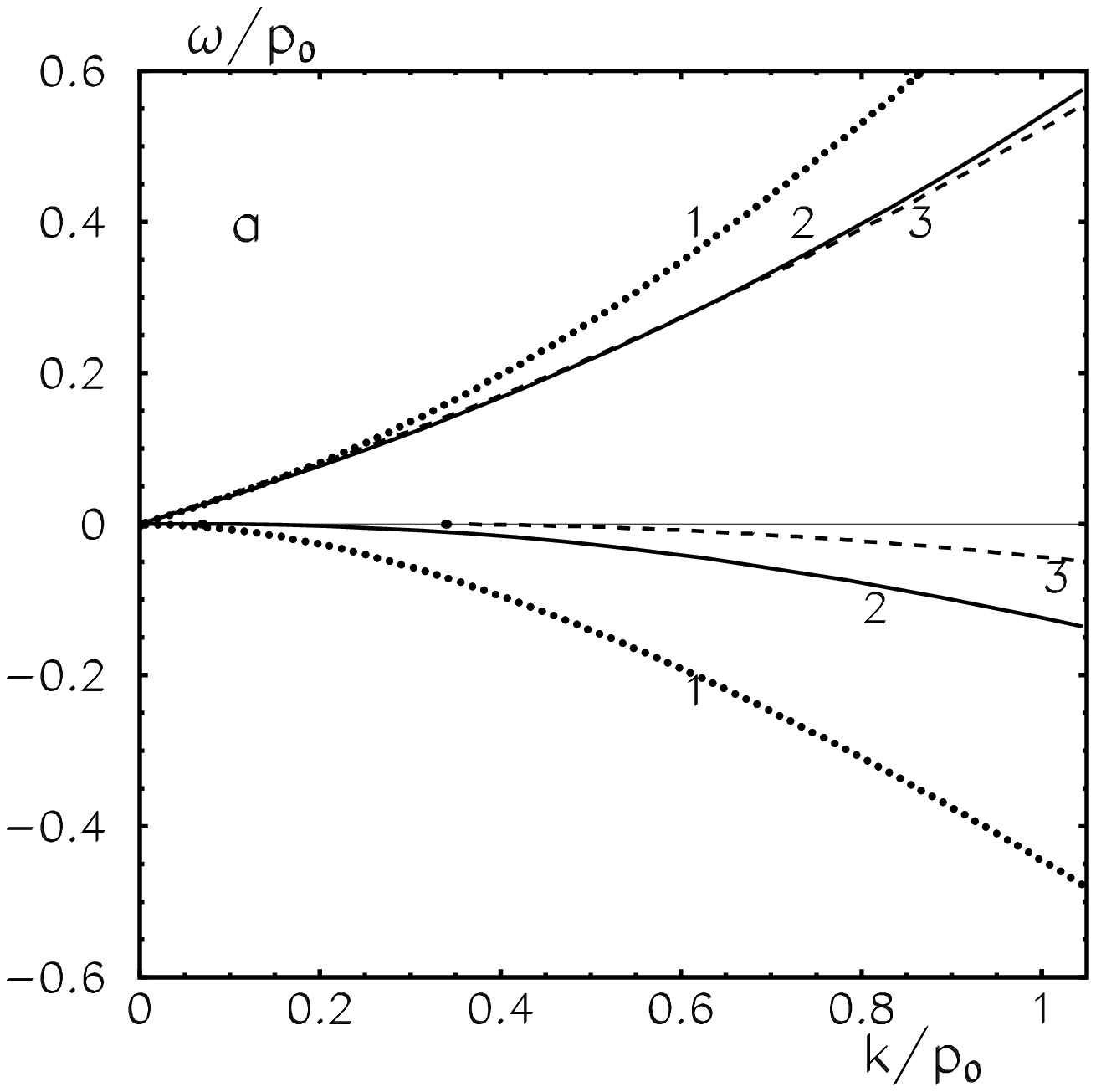,width=7cm}\epsfig{figure=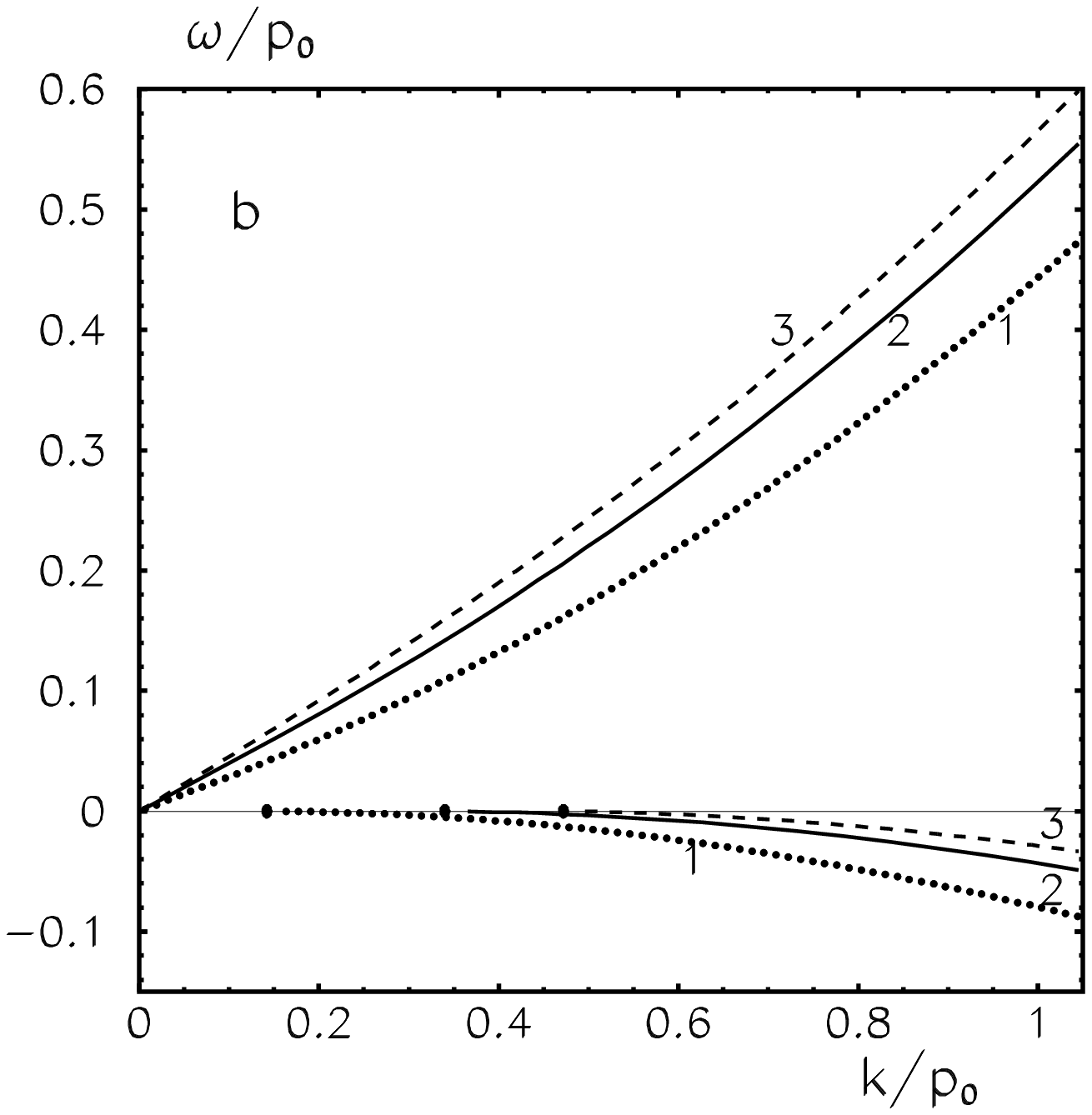,width=7cm}}
\centering{\epsfig{figure=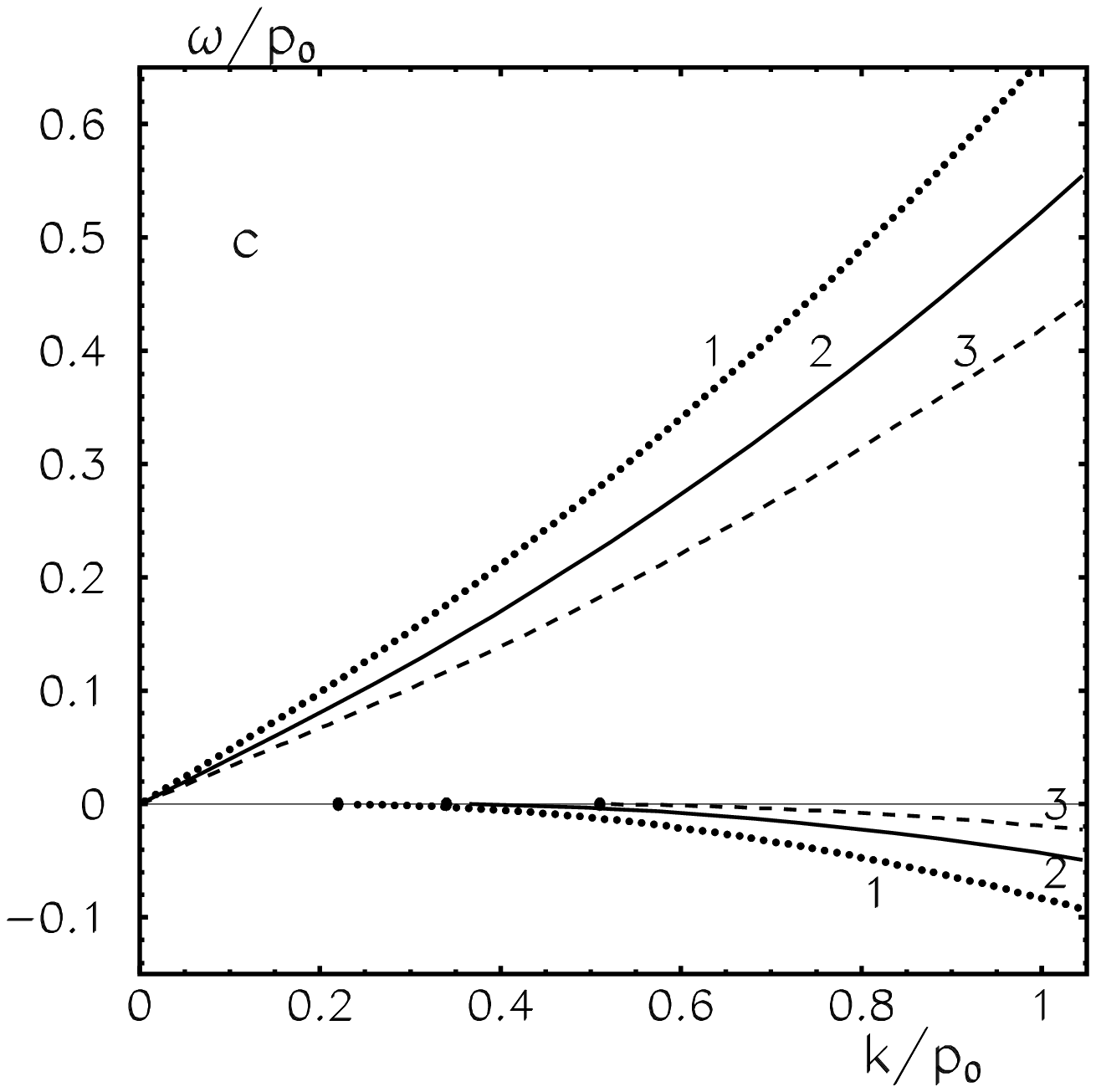,width=7cm}}
\caption{}
\end{figure}

\begin{figure}
\centering{\epsfig{figure=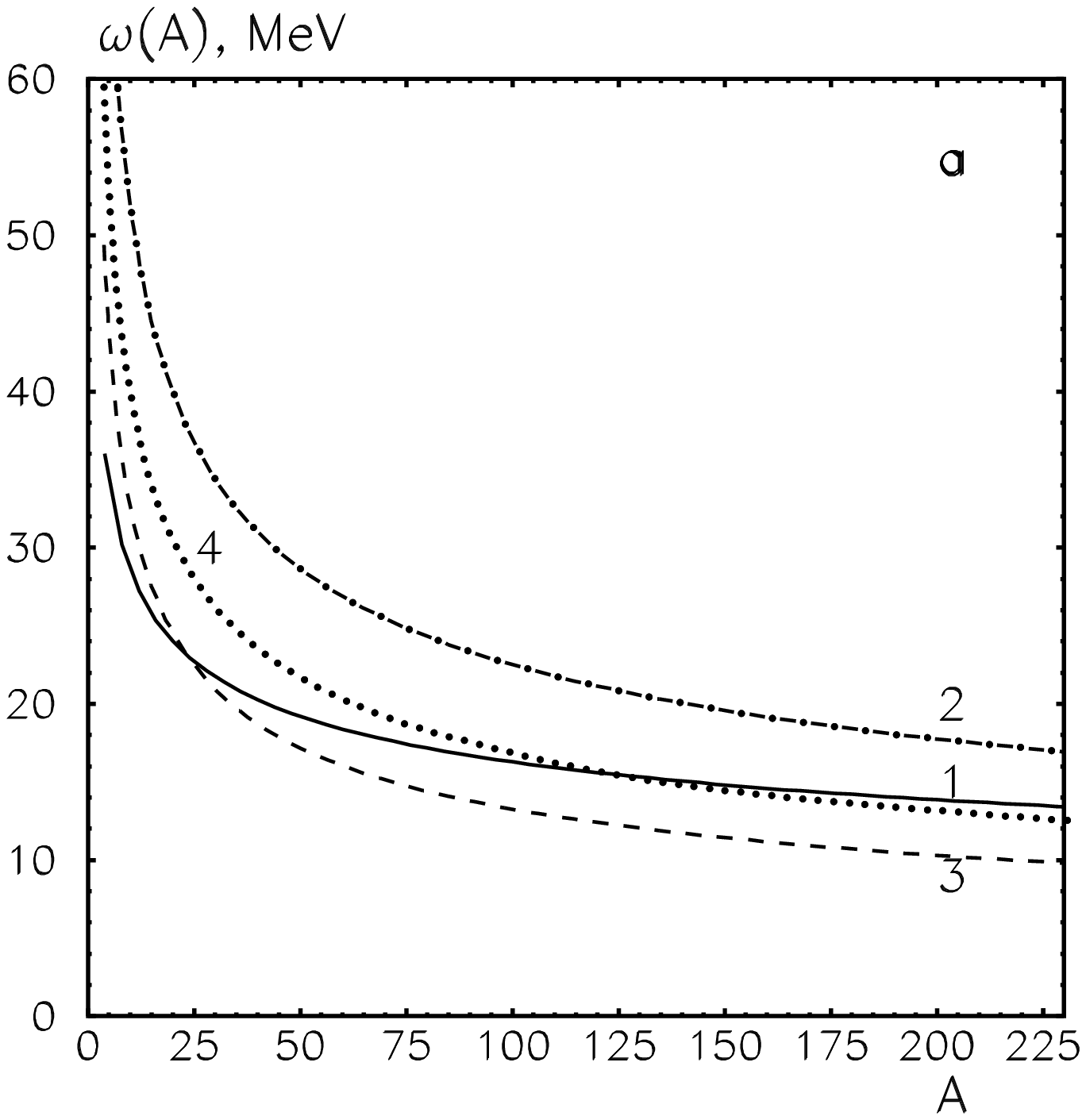,width=7cm} \epsfig{figure=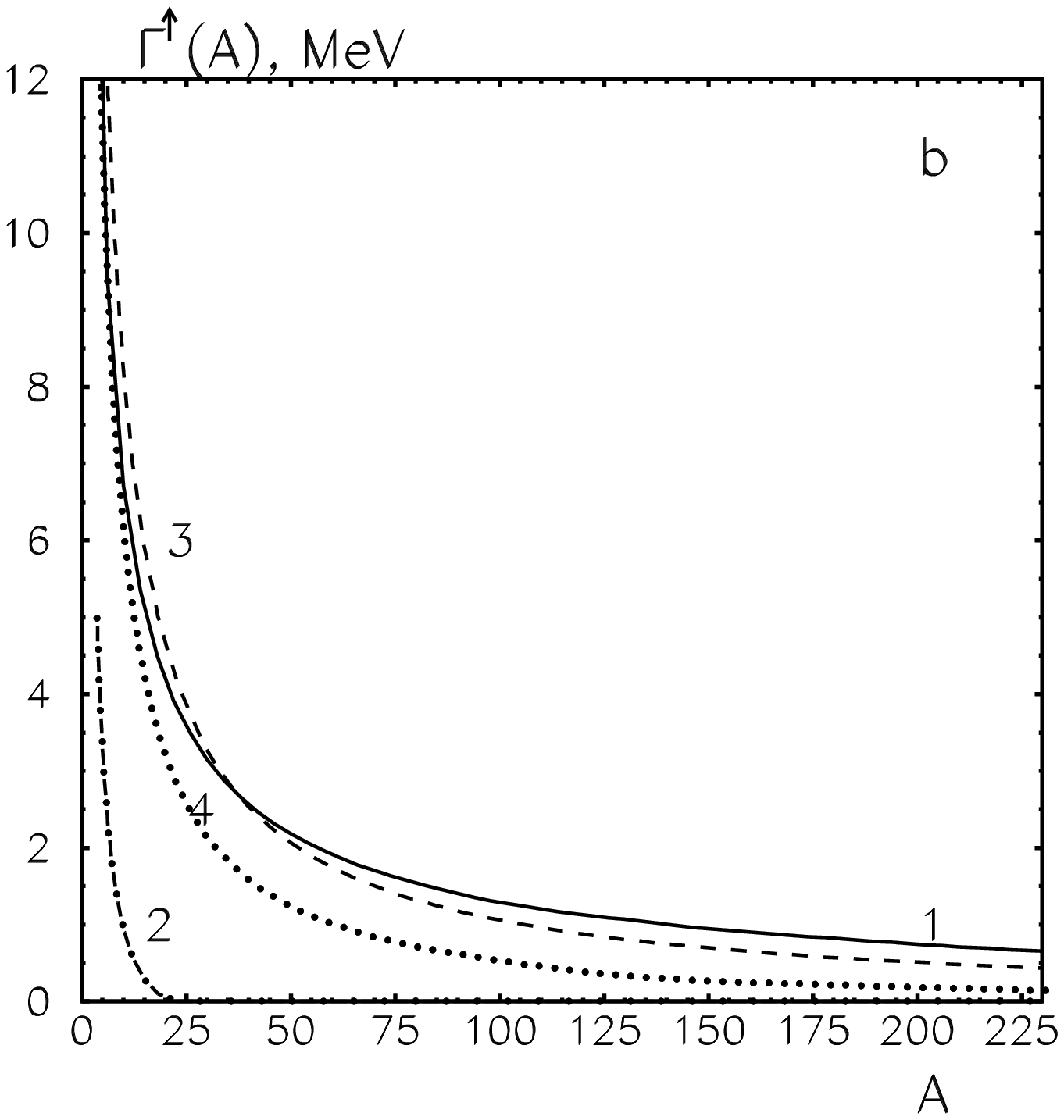,width=7cm}}
\caption{}
\end{figure}


\begin{thebibliography}{**}
\bibitem{Mi} A. B. Migdal, Theory of Finite  Fermi-Systems and Properties of the Atomic Nucleus (Willey, New-York,1967; 2 Ed. Nauka, Moskow, 1983); Rev. Mod. Phys., {\bf 50}, 107 (1078);
  A.~B.~Migdal, D.~N.~Voskresenskii, E.~E.~Saperstein, and
M.~A.~Troitskii, Phys. Rep. {\bf 192}, 179 (1990).

\bibitem{MZL} A.B. Migdal, D.F. Zaretsky, A.A. Lushnikov, Nucl. Phys., {\bf 66}, 193 (1965).

\bibitem{SW} J. Speth, J. Wambach in book "Electric and magnetic giant resonances in nuclei",
Ed. by J. Speth, 1991, World Sientific Publishing Company.
\bibitem{I31} B.S. Ishkhanov, N. P. Yudin, R. A. Eramzhyan, PEPAN, {\bf 31}, 313 (2000); B.S. Ishkhanov, I.M. Kapitonov, V.G. Neudachin, N. P. Yudin, PEPAN, {\bf 31}, 1342 (2000).
\bibitem{Ur} M. L. Gorelik, S. Shlomo, M. H. Urin, Phys.Rev., {\bf C62}, 044301  (2000).
\bibitem{De2} V.Yu. Denisov,  Phys. Atom. Nucl., {\bf 43}, 46 (1986).
\bibitem{PN} D. Pines and P. Nozieres,
{\em``The Theory of Quantum Liquids"} (W.A.Benjamin,Inc., 1966).
\bibitem{2}  V.~A.~Sadovnikova, Phys. Atom. Nucl., {\bf 70}, 989 (2007).
\bibitem{3}  V.~A.~Sadovnikova, Bull of RAS, ser. Phys., {\bf 75}, 905 (2011); arxiv.org: 1003.0547.
\bibitem{Sp} J. Speth, E. Werner, W. Wild, Phys.Rep. {\bf 33}, 127 (1977);
 S. O. Backman,  G. E. Brown,  J. A. Niskanen, Phys. Rep. {\bf 124}, 1 (1985).
 
\bibitem{SJ} P. Ring, P. Schuck, {\it``The nuclear many-body Problem"},
 (Springer-Verlag, 1980).

\bibitem{BV} F.~L. Braghin, D. Vautherin, Phys. Lett. {\bf B333}, 189 (1994).
\bibitem{BV1}   F.~L. Braghin, D. Vautherin, A.~Abada,   Phys.Rev. {\bf C52}, 2504  (1995).
\bibitem{Atomic} Atomic Data and Nuclear Data Tables, {\bf 15}, 319 (1975).
\bibitem{FW} A. L. Fetter and J. D. Walecka,{\em``Quantum theory of
many-particle systems"}(Mc-Graw-Hill, New York, 1971).

\bibitem{GLZ} D.Gambacurta, U. Lombardo, W. Zuo, Phys. Atom. Nucl., {\bf 74}, 1453, (2011).
\end{thebibliography}
\end{document}